\documentclass[12 pt,twoside,aps,prd,amsmath,amssymb,
tightenlines,showpacs,showkeys,eqsecnum]{revtex4}
\usepackage{mathptmx}
\usepackage{bm}
\usepackage{graphicx}
\usepackage{hyperref}
\renewcommand{\cal}{\mathcal}

\newcommand {\pr}{\partial}

\newcommand {\cL}{\cal L}

\newcommand {\G}{\Gamma}

\newcommand {\vf}{\varphi}

\def \myfigures #1#2#3#4#5#6#7#8
{\begin{figure}[ht]
    \begin{center}
        \includegraphics[width=#2 \textwidth]{#1.eps}
        \hfill
        \includegraphics[width=#6 \textwidth]{#5.eps}
        \parbox[t]{#4\textwidth}{\caption {#3}\label{#1}}
        \hfill
        \parbox[t]{#8\textwidth}{\caption {#7}\label{#5}}
    \end{center}
\end{figure} }

\def\myfigure #1#2#3#4
{\begin{figure}[ht]\begin{center}
\includegraphics[width=#2 \textwidth]{#1.eps}
\parbox[t]{#4\textwidth}{\caption{#3}\label{#1}}
\end{center}\end{figure}}

\date{\today}
\begin{document}
\title{Scalar field in cosmology: Potential for isotropization and inflation}
\author{Yu.P. Rybakov, G.N. Shikin and Yu.A. Popov}
\affiliation{Department of Theoretical Physics\\
Peoples' Friendship University of Russia\\
117198 Moscow, Russia} \email{soliton4@mail.ru}

\author{Bijan Saha}
\affiliation{Laboratory of Information Technologies\\
Joint Institute for Nuclear Research, Dubna\\
141980 Dubna, Moscow region, Russia} \email{bijan@jinr.ru}
\homepage{http://bijansaha.narod.ru}

\begin{abstract}
The important role of scalar field in cosmology was noticed by a
number of authors. Due to the fact that the scalar field possesses
zero spin, it was basically considered in isotropic cosmological
models. If considered in an anisotropic model, the linear scalar
field does not lead to isotropization of expansion process. One
needs to introduce scalar field with nonlinear potential for the
isotropization process to take place. In this paper the general form
of scalar field potentials leading to the asymptotic isotropization
in case of Bianchi type-I cosmological model, and inflationary
regime in case of isotropic space-time is obtained. In doing so we
solved both direct and inverse problem, where by direct problem we
mean to find metric functions and scalar field for the given
potential, whereas, the inverse problem means to find the potential
and scalar field for the given metric function. The scalar field
potentials leading to the inflation and isotropization were found
both for harmonic and proper synchronic time.

\end{abstract}

\keywords{electromagnetic field, scalar field, Bianchi type I (BI)
model, photon mass}

\pacs{03.65.Pm and 04.20.Ha}

\maketitle

\bigskip

%%%%%%%%%%%%%%%%%%%%%%%%%%%%%%%%%%%%%%%%%%%%%%%%%%%%%%%%%%%%%%%%%%%%%%%%%%

\section{Introduction}

The discovery of the accelerating mode of expansion of the Unverse
has produced a large number of papers on its possible explanation
\cite{sahni,capone,campo}. As a source of late time acceleration
beside the traditional cosmological constant
\cite{einstein,einstein1,PRpadma,lambda}, a number of other
possibilities were also suggested. To name a few are quintessence
\cite{caldwell,starobinsky,zlatev,pfdenr}, Chaplygin gas
\cite{kamen,bilic,sen,chjp}, phantom type dark energy
\cite{phantom1,dabrowski}, oscillating dark energy \cite{rubano,
stein,sahaprd}, models with interaction between dark energy and dark
matter \cite{olivares,pavon}, models with tachyon matter
\cite{srivastava,shao,copeland,cardenas,shao1}, quintom matter
\cite{Feng,cai2}, models with spinor field
\cite{SBprd04,henprd,greene,kremer1,ECAA06,sahaprd06,kremer2}.

Despite the varieties of sources able to explain the late time
acceleration, the scalar field reserves a special place in this
hierarchy. And the central idea is to find the suitable potential
that allows accelerated expansion. Due to the fact that the scalar
field possesses zero spin, it was basically considered in isotropic
cosmological models. In this paper, within the scope of Bianchi
type-I cosmological model we study the role of a scalar field with
nonlinear potential leading to both isotropization and inflation.

\section{Basic equations and their general solutions}

We consider the simplest possible scalar field model within the
framework of a BI cosmological gravitational field given by the
Lagrangian density
\begin{equation}
{\cL} = \frac{R}{2\varkappa}  + \frac{1}{2} \vf_{,\eta} \vf^{,\eta}
- V(\vf),\ \label{lag}
\end{equation}
where $\vf$ stands for real scalar field, $V(\vf)$ is the potential
and $R$ is the scalar curvature. The natural units $\hbar = c = 1$
are used. The cosmological gravitational field is given by the
Bianchi type-I (BI) metric in the form
\begin{equation}
ds^2 = e^{2\alpha} dt^2 - e^{2\beta_1} dx^2 - e^{2\beta_2} dy^2 -
e^{2\beta_3} dz^2. \label{BI}
\end{equation}
In what follows we suppose that the scalar field $\vf$ and the
metric functions $\alpha, \beta_1, \beta_2, \beta_3$ depend on $t$
only. We also assume that the metric functions satisfy the harmonic
coordinate condition
\begin{equation}
\alpha = \beta_1 + \beta_2 + \beta_3, \label{cc}
\end{equation}
implying the simplest form of the equation of motion.

Written in the form
\begin{equation}
R_{\mu}^{\nu} = -\varkappa \Bigl(T_{\mu}^{\nu} - \frac{1}{2}
\delta_{\mu}^{\nu} R \Bigr), \label{ee}
\end{equation}
the Einstein equations corresponding to the metric \eqref{BI} in
account of \eqref{cc} read
\begin{subequations}
\label{BID}
\begin{eqnarray}
e^{-2\alpha} \Bigl(\ddot \alpha - \dot \alpha^2 + \dot \beta_1^2 +
\dot \beta_2^2 + + \dot \beta_3^2 \Bigr) &=&  - \varkappa
\Bigl(T_{0}^{0} - \frac{1}{2} T \Bigr),\label{00}\\
e^{-2\alpha} \ddot \beta_1 &=&  - \varkappa
\Bigl(T_{1}^{1} - \frac{1}{2} T \Bigr),\label{11}\\
e^{-2\alpha} \ddot \beta_2 &=&  - \varkappa
\Bigl(T_{2}^{2} - \frac{1}{2} T \Bigr),\label{22}\\
e^{-2\alpha} \ddot \beta_3 &=&  - \varkappa \Bigl(T_{3}^{3} -
\frac{1}{2} T \Bigr),\label{33}
\end{eqnarray}
\end{subequations}

where over dot means differentiation with respect to $t$ and
\begin{equation}
T_{\nu}^{\mu} = \vf_{,\nu}\vf^{,\mu} - \delta_\nu^\mu
\Bigl(\frac{1}{2} \vf_{,\eta} \vf^{,\eta} - V(\vf)\Bigr),
\label{tem} \end{equation} is the energy-momentum tensor of the
scalar field.

The scalar field equation corresponding to the metric \eqref{lag}
has the form
\begin{equation}
\frac{1}{\sqrt{-g}} \frac{\pr}{\pr x^\mu} \Bigl(\sqrt{-g} g^{\mu\nu}
\vf_{,\nu} \Bigr) + \frac{dV}{d\vf} = 0. \label{scf}
\end{equation}
which on account of \eqref{cc} gives
\begin{equation}
\ddot \vf = - \frac{dV}{d \vf} e^{2\alpha}. \label{sceq}
\end{equation}

For the components of energy-momentum tensor in this case we find
\begin{subequations}
\label{emtcomp}
\begin{eqnarray}
T_0^0 &=& \frac{1}{2} \dot\vf^2 e^{-2\alpha} +
V(\vf), \label{00emt}\\
T_1^1 &=& T_2^2 = T_3^3 =  -\frac{1}{2} \dot\vf^2 e^{-2\alpha} +
V(\vf)\label{123emt}
\end{eqnarray}
\end{subequations}
In view of \eqref{emtcomp} system of Einstein equations now take the
form
\begin{subequations}
\label{BIDnew}
\begin{eqnarray}
\ddot \alpha - \dot \alpha^2 + \dot \beta_1^2 + \dot \beta_2^2 +
\dot \beta_3^2  &=&  - \varkappa \Bigl( \dot\vf^2  -
V(\vf) e^{2\alpha}\Bigr),\label{00new}\\
\ddot \beta_1 &=&  \varkappa V(\vf) e^{2\alpha},\label{11new}\\
\ddot \beta_2 &=&  \varkappa V(\vf) e^{2\alpha},\label{22new}\\
\ddot \beta_3 &=&  \varkappa V(\vf) e^{2\alpha}. \label{33new}
\end{eqnarray}
\end{subequations}
From  \eqref{11new}, \eqref{22new} and \eqref{33new} follows
\begin{equation}
\ddot \beta_1 = \ddot \beta_2 =  \ddot \beta_3. \label{beta123}
\end{equation}
On account of coordinate condition \eqref{cc} we then find
\begin{equation}
\beta_1 = (\alpha + c_1 t)/3, \quad \beta_2 = (\alpha + c_2 t)/3,
\quad \beta_3 = (\alpha + c_3 t)/3, \label{betas}
\end{equation}
with $C_i$ being the integration constants, obeying
\begin{equation}
c_1 + c_2 + c_3 = 0. \label{ic}
\end{equation}
From \eqref{ic} it follows that in case of $c_1 = c_2 = c_3 = 0$ the
expansion process is isotropic for arbitrary $\alpha (t)$, while in
case of nontrivial $C_i$'s the expansion is isotropic only if
$\alpha (t) \sim t^p$ with $p > 1$ as $t \to \infty$.

Let us consider the field equations. Summation of \eqref{11new},
\eqref{22new} and \eqref{33new}, on account of coordinate condition
gives
\begin{equation}
\ddot \alpha = 3 \varkappa V(\vf) e^{2 \alpha}. \label{al2}
\end{equation}
From \eqref{00new} in view of \eqref{betas}, \eqref{ic} and
\eqref{al2} one finds
\begin{equation}
\dot \alpha^2 - N^2 = 3 \varkappa \Bigl(\frac{1}{2} \dot\vf^2 +
V(\vf) e^{2 \alpha}\Bigr), \quad N^2 = \frac{1}{6} \Bigl(c_1^2 +
c_2^2 + c_3^2\Bigr). \label{al1}
\end{equation}

Thus we have three equations \eqref{al2}, \eqref{al1} and
\eqref{sceq} for defining $\alpha$ and $\vf$.

We are now in a position where two possible strategies could be
exploited. According to the first one we seek the solutions $\alpha
(t)$ and $\vf (t)$ to the equations \eqref{al2}, \eqref{al1} and
\eqref{sceq} for a given potential $V(\vf)$. It is known as direct
problem. Following the second strategy, we invert the problem and
construct the effective potential $V(\vf)$, hence the scalar field
$\vf (t)$ by choosing the appropriate metric function $\alpha (t)$
\cite{RSPGC}. This is known as inverse problem. We study both direct
and inverse problem in the following sections.

\section{Direct problem}

In this section we study the direct problem solving the equations
\eqref{al2}, \eqref{al1}, \eqref{sceq} to define $\alpha (t)$ and
$\vf)t)$ for a given potential $V(\vf)$. Since the present day
Universe is surprisingly isotropic,  our main objective will be to
define potential that leads to isotropization of the initially
anisotropic space-time.

Let us choose the potential in the form:
\begin{equation}
V(\vf) = V_0 e^{\lambda \vf(t)}, \quad \lambda = {\rm const.}
\label{potass}
\end{equation}
The corresponding equations now take the form
\begin{equation}
\ddot \alpha = 3 \varkappa V_0 e^{\lambda \vf + 2 \alpha},
\label{al2n}
\end{equation}
\begin{equation}
\dot \alpha^2 - N^2 = 3 \varkappa \Bigl(\frac{1}{2} \dot\vf^2 + V_0
e^{\lambda \vf + 2 \alpha}\Bigr),  \label{al1n}
\end{equation}
\begin{equation}
\ddot \vf = - \lambda V_0 e^{\lambda \vf + 2\alpha}. \label{sceqn}
\end{equation}
Equations \eqref{al2n} and \eqref{sceqn} leads to
\begin{equation}
\lambda \ddot \alpha + 3 \varkappa \ddot \vf = 0, \label{vfa}
\end{equation}
with the solution
\begin{equation}
\lambda \alpha + 3 \varkappa \vf = A_1 t, \quad A_1 = {\rm const.}
\label{vfan}
\end{equation}
Inserting $\vf$ from \eqref{vfan} into \eqref{al2n} we find
\begin{equation}
\ddot \alpha = 3\varkappa V_0 e^{\frac{\alpha(6 \varkappa -
\lambda^2)}{3 \varkappa} + At}, \quad  A = \frac{\lambda A_1}{3
\varkappa}. \label{aleq}
\end{equation}
Depending on $6 \varkappa - \lambda^2$ there occurs three
possibilities.

\vskip 5 mm

{\bf Case I}

Let us consider the case with $6\varkappa - \lambda^2 = p_1^2 > 0$.
Introducing
\begin{equation}
\eta (t) = \frac{p_1^2}{3\varkappa} \alpha + At, \label{eta}
\end{equation}
from \eqref{aleq} one finds
\begin{equation}
\ddot \eta = p_1^2 V_0 e^\eta. \label{eta1}
\end{equation}
Solving the Eq. \eqref{eta1} we obtain
\begin{equation}
e^{\eta} = \frac{c_4}{2p_1^2 V_0} \frac{1}{\sinh{(\sqrt{c_4}t/2)}}.
\label{eta10}
\end{equation}

Inserting \eqref{eta10} into \eqref{eta} we find
\begin{equation}
\alpha (t) = \frac{3\varkappa}{p_1^2} \Bigl[\ln{\bigl(c_4/2p_1^2
V_0\bigr)} - 2 \ln\bigl[{\sinh{\bigl(\sqrt{c_4} t/2\bigr)}}\bigr] -
At\Bigr]. \label{alf1}
\end{equation}
As one sees, $\alpha$ is a linear function of $t$. Thus we conclude
that in case of $6\varkappa - \lambda^2 = p_1^2 > 0$ the metric
coefficients are the linear functions of $t$ and in this case no
isotropization process takes place.

\vskip 5 mm

{\bf Case II}

Let us consider the case with $6\varkappa - \lambda^2 = - p_2^2 <
0$. Introducing
\begin{equation}
\xi (t) = -\frac{p_2^2}{3\varkappa} \alpha + At, \label{xi}
\end{equation}
from \eqref{aleq} one finds
\begin{equation}
\ddot \xi = -p_2^2 V_0 e^\xi. \label{xi1}
\end{equation}
The Eq. \eqref{xi1} has the following solution
\begin{equation}
e^{\xi} = \frac{c_4}{2p_2^2 V_0} \frac{1}{\cosh{(\sqrt{c_4}t/2)}}.
\label{xi10}
\end{equation}

Inserting \eqref{xi10} into \eqref{xi} one gets
\begin{equation}
\alpha (t) = -\frac{3\varkappa}{p_2^2} \Bigl[\ln{\bigl(c_4/2p_2^2
V_0\bigr)} - 2 \ln\bigl[{\cosh{\bigl(\sqrt{c_4} t/2\bigr)}}\bigr] -
At\Bigr]. \label{alf2}
\end{equation}
As one sees, $\alpha$ is a linear function of $t$. In means, as in
case I, the assumption $6\varkappa - \lambda^2 = - p_2^2 < 0$ leads
to the metric coefficients be the linear functions of $t$ and in the
case in hand no isotropization process takes place.

\vskip 5 mm

{\bf Case III}

Let us consider the case with $(6\varkappa - \lambda^2) =  0$. We
will study the partial case setting $A_1 = 0$. Then for $\alpha$ and
$\vf$ we have
\begin{equation}
\ddot \alpha = 3 \varkappa V_0, \quad \Rightarrow \quad \alpha =
\frac{3\varkappa V_0}{2} t^2 + c_5 t + c_{50}, \label{al0}
\end{equation}
\begin{equation}
\ddot \vf = -\lambda V_0, \quad \Rightarrow  \quad \vf =
-\frac{\lambda V_0}{2} t^2 + c_6 t + c_{60}. \label{vf0}
\end{equation}

Thus we see that for $\lambda = \pm \sqrt{6\varkappa}$, $\alpha$ is
a quadratic function of $t$, which allows isotropic expansion of the
Universe. This result is in agreement with \eqref{potsim}. Hence,
for $n = 0$, only the potential of type \eqref{potsim} leads to
isotropization.

\section{Inverse problem}

In this section we study the inverse problem. Here we seek the
potential for the given metric functions $\alpha$ and scalar field
$\vf$. As some special cases we look for the potentials leading to
the isotropization of initially anisotropic metric and inflationary
scenario.

Let us first study the inverse problem in general. In doing so we
assume
\begin{equation}
\alpha (t) = \G[\vf(t)]. \label{assume}
\end{equation}
Inserting \eqref{assume} into \eqref{al1} we find
\begin{equation}
\dot \vf^2 = \frac{N^2 + 3 \varkappa V(\vf) e^{2\G}}{\G_{\vf}^2
 - 3 \varkappa/2}, \quad \G_{\vf} = \frac{d\G}{d\vf}. \label{fi2}
\end{equation}
Differentiation of \eqref{fi2} with respect to $t$ on account of
\eqref{sceq} gives
\begin{equation}
V_\vf + M(\vf) V + H(\vf) = 0, \label{neq}
\end{equation}
with
\begin{equation}
M(\vf) = \frac{3 \varkappa}{\G_\vf}\Biggl(1 -
\frac{\G_{\vf\vf}}{\G_{\vf}^2 - 3 \varkappa/2}\Biggr), \quad H(\vf)
= - N^2 \frac{e^{-2\G}}{\G_{\vf}^2  - 3 \varkappa/2}
\frac{\G_{\vf\vf}}{\G_{\vf}}. \label{neq1}
\end{equation}
The equation \eqref{neq} is the first order linear inhomogeneous
equation with the general solution
\begin{equation}
V(\vf) = e^{-\int M d\vf} \Biggl[c_4 - \int d\vf H(\vf)e^{\int M
d\vf} \Biggl], \quad c_4 = {\rm const.}, \label{potsol}
\end{equation}
which can be viewed as equation for defining $V(\vf)$ for a given
metric function $\G(\vf)$. In what follows we find the potentials
$V(\vf)$ leading to some specific cosmological solutions.

\subsection{Potential for isotropization}

In this subsection we looking for the potentials leading to the
isotropization of initially anisotropic metric. In doing so we
consider two cases.

\subsubsection{Asymptotically isotropic space-time}

Let us find potential that leads to the asymptotic isotropization of
the initially anisotropic model. Note that  the space-time becomes
asymptotically isotropic if:
\begin{equation}
\alpha (t)\bigl|_{t \to \infty} \propto \alpha_0 t^p, \quad p
> 1, \quad \alpha_0 = {\rm const}. \label{ispot}
\end{equation}
Inserting $\alpha (t)$ from \eqref{ispot} into \eqref{al2} one finds
\begin{equation}
\alpha_0 p(p-1)t^{2p-2} = 3 \varkappa V(\vf) e^{2\alpha_0 t^p}.
\label{al3}
\end{equation}
Further inserting $\alpha (t)$ into \eqref{al1} on account of
\eqref{al3} we have
\begin{equation}
\alpha_0^2 p^2 t^{2p-2} - N^2 = \frac{3 \varkappa}{2} \dot \vf^2 +
\alpha_0 p(p-1)t^{p-2}.\label{fi01}
\end{equation}
From \eqref{fi01} one finds
\begin{equation}
\vf (t)|_{t \to \infty} = \pm \alpha_0 \sqrt{\frac{2}{3\varkappa}}
t^p = \pm \sqrt{\frac{2}{3\varkappa}} \alpha (t). \label{alfi}
\end{equation}
From \eqref{alfi} one finds
\begin{equation}
t = \Bigl(\pm \frac{1}{\alpha_0}\sqrt{\frac{3\varkappa}{2}}
\vf(t)\Bigr)^{1/p}. \label{time}
\end{equation}
Inserting $t$ from \eqref{time} into \eqref{al3} one finds
\begin{equation}
3 \varkappa V(\vf) = \alpha_0 p(p-1)\Bigl(\pm
\frac{1}{\alpha_0}\sqrt{\frac{3\varkappa}{2}} \vf(t)\Bigr)^{(p-2)/p}
e^{\pm2\sqrt{\frac{3\varkappa}{2}} \vf(t)}. \label{pot}
\end{equation}
The general form of potential, leading to isptropization, is:
\begin{equation}
V(\vf) = V_0 \vf^n e^{\sqrt{6\varkappa} \vf(t)}, \quad n =
\frac{p-2}{p}, \quad p > 1, \quad -1 < n < 1. \label{potgen}
\end{equation}
The simplest form of scalar field potential is obtained for $p = 2$,
$n = 0$:
\begin{equation}
V(\vf) = V_0 e^{\sqrt{6\varkappa} \vf(t)}.  \label{potsim}
\end{equation}
Thus we find that the potential given by \eqref{potsim} leads to the
asymptotic isotropization of initially anisotropic space-time.

\subsubsection{Isotropic space-time}

Let us consider the isotropic model when $\beta_1 (t) = \beta_2 (t)
= \beta_3 (t) = \alpha(t)/3$, i.e., $c_1 = c_2 = c_3 = 0$ in
\eqref{betas}. It means $N^2 = 0$ in \eqref{al1} and $H(\vf) = 0$ in
\eqref{neq}. The metric and basic system of equations in this case
takes the form:

\begin{equation}
ds^2 = e^{2\alpha} dt^2 - e^{2\alpha/3}\bigl(dx^2 + dy^2 +
dz^2\bigr). \label{BIi}
\end{equation}
\begin{equation}
\ddot \alpha = 3 \varkappa V(\vf) e^{2\alpha}, \label{al1i}
\end{equation}
\begin{equation}
\dot \alpha^2 = 3 \varkappa \bigl(\frac{1}{2} \dot \vf^2 + V(\vf)
e^{2\alpha}\bigr), \label{al2i}
\end{equation}
\begin{equation}
\ddot \vf = -\frac{d V(\vf)}{d\vf} e^{2\alpha}. \label{sci}
\end{equation}
In this case from \eqref{neq} we find
\begin{equation}
V(\vf) = c_4 \Bigl(\frac{3\varkappa}{2} \frac{1}{\Gamma_\vf^2} -
1\Bigr)e^{-3\varkappa \int \frac{d\vf}{\Gamma_\vf}}, \quad
\Gamma(\vf) = \alpha(t), \quad \Gamma_\vf = \frac{d\Gamma}{d\vf}.
\label{coni}
\end{equation}

For $\Gamma(\vf)$ from \eqref{neq} we find
\begin{equation}
\Gamma_{\vf\vf} - \frac{1}{3\varkappa} \frac{V_\vf}{V}\Gamma_\vf^3 -
\Gamma_\vf^2 + \frac{1}{2}\frac{V_\vf}{V}\Gamma_\vf +
\frac{3\varkappa}{2} = 0. \label{Gam}
\end{equation}
It is the Abel equation of first kind.

Introducing a new function
\begin{equation}
U(\psi) = e^{-\frac{3\varkappa}{2} \int\frac{d\vf}{\Gamma_\vf}},
\quad \psi = \sqrt{3\varkappa}\, \vf, \label{NF}
\end{equation}
from \eqref{coni} we find
\begin{equation}
\frac{V}{c_4} = U^{\prime 2} - U^2, \quad U^{\prime} =
\frac{dU}{d\psi}. \label{VU}
\end{equation}
The equation \eqref{VU} possesses exact solution for some $V(\vf)$.
Some simple cases are
\begin{eqnarray}
V &=& c_4 \cos{2\psi} \qquad \Rightarrow \qquad   U = \sin{\psi}, \\
V &=& 0 \qquad  \Rightarrow \qquad   U = \exp{\psi}, \\
V &=& c_4 \qquad \Rightarrow \qquad   U = \sinh{\psi}.
\end{eqnarray}

\subsection{Potential for inflation}

Let us define the scalar field potential leading to inflation. The
isotropic model of the Universe in harmonic time is given by the
metric
\begin{equation}
ds^2 = e^{2\alpha (t)} dt^2 - e^{\frac{2}{3}\alpha (t)} \bigl(dx^2 +
dy^2 + dz^2\bigr), \label{methar}
\end{equation}
which in proper synchronic time takes the form
\begin{equation}
ds^2 =  d\tau^2 - a^2 (\tau) \bigl(dx^2 + dy^2 + dz^2\bigr).
\label{metsyn}
\end{equation}
In inflation regime $a(\tau) \sim t^p$,\,\, $p > 1$. Comparing
\eqref{methar} and \eqref{metsyn} one finds
\begin{eqnarray}
d\tau &=& \pm e^{\alpha (t)} dt, \quad \Rightarrow \quad \dot \tau =
\frac{d\tau}{dt} = \pm e^{\alpha (t)}, \quad \Rightarrow \quad \tau
= \pm \int e^{\alpha (t)} dt, \label{dtau}\\
e^{\frac{1}{3}\alpha (t)} &=& a(\tau), \quad \Rightarrow \quad
\bigl(\dot \tau\bigr)^{1/3} = a(\tau),\quad \Rightarrow \quad \int
[a(\tau)]^{-3} d\tau = t + t_0. \label{atau}
\end{eqnarray}
For inflation regime the function $\tau(t)$ is determined from
\eqref{atau} for the given $a(\tau)$. From \eqref{dtau} we have $\pm
\alpha (t) = \ln{\dot \tau}$. Defining $\dot \tau$ for the given
$a(\tau)$ from \eqref{atau} one finds $\alpha(t)$. Further inserting
$\alpha (t)$ into \eqref{al1i} and \eqref{al2i} we find $\vf(t)$ and
$V(\vf)$, i.e., potential that leads to inflation. In this regime
\begin{equation}
a (\tau) = (H\tau)^p, \quad H = {\rm const.}, \quad p = {\rm const.}
> 1. \label{ainf}
\end{equation}
In this case from \eqref{atau} follows
\begin{equation}
\int (H\tau)^{-3p} d\tau = \frac{1}{H} \frac{(H\tau)^{1-3p}}{1-3p} =
t, \label{Htau}
\end{equation}
that gives
\begin{equation}
\tau(t) = \frac{1}{H} (AHt)^{1/A}, \quad e^{\alpha (t)}= \dot \tau =
(AHt)^B, \quad A = 1 - 3p < 0, \quad B = \frac{1}{A} - 1.
\label{taut}
\end{equation}
After a little manipulation one finds
\begin{equation}
\vf(t) = \pm {\cal D}  \ln{|t|}, \quad {\cal D} =
\sqrt{\frac{2}{3\varkappa}B(B+1)}, \label{scinf}
\end{equation}
and
\begin{equation}
V(\vf) = V_0 e^{\pm \sqrt{2\varkappa/p}\,\vf}. \label{potinf}
\end{equation}
Thus we found $V(\vf)$ in harmonic coordinates, leading to the power
law expansion in proper synchronic time. $V(\vf) \to {\rm const.}$
as $p \to \infty$, which corresponds to the cosmological constant.

\subsection{Power law expansion}

Finally, we define the scalar field potential leading to the power
law expansion in the proper synchronic time. The initial metric in
this case takes the form \eqref{metsyn}. In this case $a(\tau)$ is
given by \eqref{ainf} with $p$ being arbitrary constant. The
Einstein equation for \eqref{metsyn} takes the form
\begin{equation}
3\frac{\dot a^2}{a^2} = \varkappa T_0^0 = \varkappa
\bigl[\frac{1}{2} \dot \vf^2 + V(\vf)\bigr], \label{00sin}
\end{equation}
\begin{equation}
2\frac{\ddot a}{a} + \frac{\dot a^2}{a^2} = \varkappa T_i^i =
\varkappa \bigl[-\frac{1}{2} \dot \vf^2 + V(\vf)\bigr].
\label{iisin}
\end{equation}
Subtraction \eqref{iisin} from \eqref{00sin} on account \eqref{ainf}
gives
\begin{equation}
\vf = \pm \sqrt{\frac{2p}{\varkappa}} \bigl(\ln{\tau} +
\ln{\tau_0}\bigr), \quad \Rightarrow \quad \tau = \tau_0 e^{\pm
\sqrt{\frac{\varkappa}{2p}} \vf}. \label{scsyn}
\end{equation}
Summation of \eqref{iisin} and \eqref{00sin} in this case gives
\begin{equation}
V(\vf) = V_0 e^{\pm \sqrt{\frac{2\varkappa}{p}} \vf}, \quad V_0 =
\frac{p(3p-1)}{\varkappa \tau_0^2}, \label{potsyn}
\end{equation}
which coincides with that of \eqref{potinf}.

\section{Conclusion}

The general form of scalar field potentials leading to the
asymptotic isotropization in case of  Bianchi type-I cosmological
model, and inflationary regime in case of isotropic space-time is
obtained. In doing so we solved both direct and inverse problem. The
scalar field potentials leading to the inflation and isotropization
were found both for harmonic and  proper synchronic time. It is
shown that in the cases considered, the potential that leads to
isotropization  takes the form $V(\vf) = V_0 \vf^n e^{\pm
\sqrt{6\varkappa} \vf}$, with $n \in (-1,\,+1)$ being some constant.
In case of $n=0$ we find $V(\vf) = V_0 e^{\pm \sqrt{6\varkappa}
\vf}$. The last one leads to inflation as well.

\end{document}